\documentstyle[12pt]{article}
\newlength{\dinwidth}
\newlength{\dinmargin}
\def\lapproxeq{\lower .7ex\hbox{$\;\stackrel{\textstyle <}{\sim}\;$}}
\def\gapproxeq{\lower .7ex\hbox{$\;\stackrel{\textstyle >}{\sim}\;$}}
\def\pp{{ p\bar{p}}}
\def\tt{t\bar{t}}
\def\nb{{\rm nb}}
\def\pb{{\rm pb}}
\def\lmsb{\Lambda_{\overline{\rm MS}}}
\def\MeV{{\rm MeV}}
\def\GeV{{\rm GeV}}
\def\TeV{{\rm TeV}}
\def\btop{\mathchar"1339}
\def\bmid{\mathchar"133D}
\def\bbot{\mathchar"133B}

\begin{document}
\begin{titlepage}
\vspace*{-1cm}
\begin{flushright}
RAL-94-055 \\
DTP/94/34   \\
June  1994\\
\end{flushright}
\vskip 1.cm
\begin{center}
{\Large\bf Parton Distributions of the Proton}
\vskip 1.cm
{\large A.D.~Martin} and {\large W.J.~Stirling}
\vskip .2cm
{\it Department of Physics, University of Durham \\
Durham DH1 3LE, England }\\
\vskip .4cm
and
\vskip   .4cm
{\large R.G.~Roberts}
\vskip .2cm
{\it
Rutherford Appleton Laboratory,  \\
Chilton, Didcot  OX11 0QX, England
} \\
\vskip 1cm
\end{center}
\begin{abstract}
To obtain improved parton densities of the proton, we present a new
global analysis of deep inelastic and related data including, in
particular, the recent measurements of $F_2$ at HERA, of the asymmetry
of the rapidity distributions of $W^\pm$ production at the FNAL $\pp$
collider and of the asymmetry in Drell-Yan production in $pp$ and $pn$
collisions. We  also incorporate data to determine the flavour
dependence of the quark sea distributions.
We find that the behaviour of the partons at small $x$ is
consistent with the precocious onset of BFKL leading $\log(1/x)$
dynamics. We discuss the ambiguities remaining in the gluon
distribution. We present improved predictions for $W$ boson (and $t$
quark)  production at the FNAL $\pp$ collider.
\end{abstract}
\vfill
\end{titlepage}

\newpage

\noindent {\large {\bf 1.  Introduction}}

The increased precision in the experimental measurements of deep inelastic
scattering and related processes over the last few years has led to a
considerable improvement in our knowledge of the parton distributions of the
proton.  However several significant sets of measurements have become available
since the last global analyses of the data were performed to determine the
parton densities.  The new data may be divided into two groups.  First we have
the measurements of the structure function $F_2(x,Q^2)$ for electron-proton
deep-inelastic scattering for $x \lapproxeq 10^{-3}$ by the H1 and ZEUS
collaborations at HERA \cite{HERA,H1,ZEUS}, and secondly the measurement of the
asymmetry in Drell-Yan production in $pp$ and $pn$ collisions by the NA51
collaboration at CERN~\cite{NA51}, and of the asymmetry of the $W^{\pm}$
rapidity distributions by the CDF collaboration at FNAL~\cite{BODEK}.  The two
groups of measurements have quite distinct implications for the partons and
are,
therefore to a great extent, decoupled.  The HERA $F_2$ data offer the first
constraints on partons in the previously unexplored small $x$ regime, whereas
the two asymmetry measurements probe fine details of the quark distributions in
the region $x \sim 0.1$.  The latter information is crucial, for example, for a
precise determination of the mass of the $W$ boson at the FNAL collider.

Fig.~1 shows the dramatic rise of $F_2$ with decreasing $x$, which was first
observed by the H1 and ZEUS collaborations \cite{HERA}.  Also shown are the
extrapolated predictions for $F_2$ at $Q^2 = 15$ GeV$^2$ from two parton sets
(D$^{\prime}_0$,D$^{\prime}_-)$ \cite{MRSD} made before the HERA measurements
became available, but which it was believed would span the data.  The upper
limit, the curve D$^{\prime}_-$, was motivated by assuming the precocious onset
of BFKL dynamics \cite{BFKL} in which the gluon and sea quark distributions
have
the singular form
\begin{equation}
xg, \; xq_{{\rm sea}} \; \sim \; x^{-0.5}
\end{equation}
as $x \rightarrow 0$, whereas the D$^{\prime}_0$ curve was the lower limit
anticipated from conventional Regge expectations with
\begin{equation}
xg, \; xq_{{\rm sea}} \; \sim \; x^0 .
\end{equation}
Such extrapolations are notoriously unreliable and have failed in the past.
Moreover, as we will see in section 3, the connection with precocious BFKL
behaviour turns out to be more subtle than (1) would suggest.  Although
D$^{\prime}_0$ and D$^{\prime}_-$ are the most recent published MRS sets of
partons, a set MRS(H) was subsequently made available,\footnote{A brief
description of MRS(H) can be found in Ref.~\cite{MRSH}.} which was obtained
{}from a global analysis that incorporated the measurements \cite{HERA} of
$F_2$
{}from the 1992 HERA run.  The curve denoted by H on Fig.~1 is an example of
the quality of the fit.  The main feature is that the HERA measurements
required
a small $x$ behaviour of the form
\begin{equation}
xg, \; xq_{{\rm sea}} \; \sim \; x^{-0.3} .
\end{equation}
Higher statistics measurements of $F_2$, obtained from the 1993 HERA run, have
just become available in preliminary form \cite{H1,ZEUS}.  We incorporate these
data in our new global analysis and in section 3 we discuss the implications
for
the small $x$ behaviour of the partons and for QCD dynamics.  The new HERA data
for $F_2$ are in line with the old (i.e.\ show the same rise with decreasing
$x$), and in fact the MRS(H) partons still give an excellent fit in the HERA
small $x$ region.

The data in Figs.~2 and 3 are respectively the measurement of the asymmetry in
Drell-Yan production in $pp$ and $pn$ collisions \cite{NA51}
\begin{equation}
A_{DY} \; = \; \frac{\sigma_{pp}-\sigma_{pn}}{\sigma_{pp}+\sigma_{pn}} ,
\label{eq:ady}
\end{equation}
and of the asymmetry of the rapidity distributions of the charged leptons from
$W^{\pm} \rightarrow \ell^{\pm}\nu$ decays at FNAL \cite{BODEK},
\begin{equation}
A(y_{\ell}) \; = \;
\frac{\sigma(\ell^+)-\sigma(\ell^-)}{\sigma(\ell^+)+\sigma(\ell^-)} .
\label{lepasy}
\end{equation}
In (\ref{eq:ady}), $\sigma \equiv d^2\sigma/dMdy|_{y=0}$ where $M$ and $y$ are
the
invariant mass and rapidity of the produced lepton-pair, while in
(\ref{lepasy}) $\sigma (\ell^\pm) \equiv d\sigma/dy_{\ell}$
are the differential $\pp\to W^\pm X \to \ell^\pm \nu X$ cross sections
for producing $\ell^\pm$ leptons of rapidity $y_\ell$.
  Also shown are the predictions of MRS(H) and the equivalent set of
partons, CTEQ2M, obtained by the CTEQ collaboration \cite{CTEQ}.  We see that
neither set of partons gives a satisfactory description of both asymmetries.
This deficiency of the parton sets is not surprising.  The reason is that the
high-precision muon and neutrino deep-inelastic structure function data, which
provide the core constraints of the global analyses, do not pin down the
combination $\bar{d}-\bar{u}$ of parton densities.  Indeed the Drell-Yan
asymmetry experiment was proposed \cite{ES} as it was uniquely equipped to
determine just this combination of densities.  The asymmetry data therefore
offer a fine-tuning of the $u,d,\bar{u}$ and $\bar{d}$ parton densities in the
region $x \sim 0.1$, which is invaluable for the precision studies of the $W$
boson at FNAL.  To this end we include for the first time the asymmetry data in
the global analysis (together with the new HERA measurements of $F_2$) and find
a new set of partons, which we denote MRS(A).  The resulting description of the
asymmetry measurements is also shown in Figs.~2 and 3.  We discuss this aspect
of the global analysis in section 4.

The outline of the paper is as follows.  We first explain, in section 2, the
procedure that we follow to determine the parton densities from a global
analysis of the data. We give details of the new
improved MRS(A) parton distributions, and we compare them with the MRS(H) set.
Sections 3 and 4 consider the impact of the new small $x$
and asymmetry data respectively.
 In section 5 we discuss the ambiguities in the present knowledge of the gluon.
In section 6 we update the predictions for $W$ boson and top quark production
at the FNAL $\pp$ collider, and finally in section 7 we present our
conclusions.

\vspace*{1cm}

\noindent {\large {\bf 2.  The global analysis}}

The parton distributions $f_i$ are determined from a global fit to a wide range
of deep-inelastic and related data.  The basic procedure is to parametrize the
$f_i$ at a sufficiently large $Q^2_0$ ($Q^2_0 = 4$ GeV$^2$) so that
$f_i(x,Q^2)$
can be calculated at higher $Q^2$ in perturbative QCD using next-to-leading
order Altarelli-Parisi (GLAP) evolution equations.  In view of the quantity and
variety of data that are fitted, it is remarkable that an excellent description
can be obtained with the following simple parametrization
\begin{eqnarray}
xu_{\rm v} & = & A_u x^{\eta_1}(1-x)^{\eta_2} (1 + \epsilon_u \sqrt{x} +
\gamma_u x) \nonumber \\
xd_{\rm v} & = & A_d x^{\eta_3}(1-x)^{\eta_4} (1 + \epsilon_d \sqrt{x} +
\gamma_dx) \nonumber \\
xS & = & A_S x^{-\lambda}(1-x)^{\eta_S}(1 + \epsilon_S \sqrt{x} + \gamma_Sx)
\nonumber \\
xg & = & A_g x^{-\lambda}(1-x)^{\eta_g} (1 + \gamma_gx) ,
\label{eq:starting}
\end{eqnarray}
where the valence distributions $u_{\rm v} \equiv u-\bar{u}$ and $d_{\rm v}
\equiv d-\bar{d}$, and where the total  sea distribution $S \equiv
2(\bar{u}+\bar{d}+\bar{s}+\bar{c})$.  We assume that $s = \bar{s}$.
 At present there
are not enough experimental constraints on the gluon to justify the
introduction
of an extra parameter $\epsilon_g$ in $xg$,
or to determine the exponent $\lambda$ independent of that of the sea-quark
distribution $S$.
  Three of the four $A_i$
coefficients are determined by the momentum and flavour sum rules.  The
distributions are defined in the $\overline{{\rm MS}}$ renormalization and
factorization scheme and the QCD scale parameter $\Lambda_{\overline{{\rm
MS}}}(n_f = 4)$ is taken as a free parameter.

The flavour structure of the quark sea is taken to be
\begin{eqnarray}
2\bar{u} & = & 0.4 (1-\delta) S - \Delta \nonumber \\
2\bar{d} & = & 0.4 (1-\delta) S + \Delta \nonumber  \\
2\bar{s} & = & 0.2 (1-\delta) S \nonumber  \\
2\bar{c} & = & \delta S
\label{eq:sea}
\end{eqnarray}
at $Q^2 = Q_0^2 = 4\ \GeV^2$, with
\begin{equation}
x\Delta \; \equiv \; x(\bar{d}-\bar{u}) \; = \; A_{\Delta} x^{\eta_1}
(1-x)^{\eta_S} (1 + \gamma_{\Delta}x) .
\label{eq:delta}
\end{equation}
The first hint that the $u,d$  flavour symmetry of the  sea is broken (with
$\bar{d} > \bar{u}$ on average) came from the evaluation of the Gottfried sum
by
NMC \cite{NMC}.  In fact until then all global analyses had assumed $\bar{u} =
\bar{d}$.  As we shall see in section 4, the observed Drell-Yan asymmetry
provides further evidence that $\bar{d} > \bar{u}$, which we allow through the
parametrization of $\Delta$ given in (\ref{eq:delta}).  The flavour breaking
can be
associated with the breaking of $\rho-a_2$ meson Regge exchange degeneracy and
so we choose the exponent $\eta_1$ in (\ref{eq:delta}) to be the same as that
of one of the
valence quark densities.

As in earlier MRS parametrizations, we assume that the input strange
sea is suppressed by 50\% in relation to the $u$ and $d$ sea distributions --
hence the factors 0.4, 0.4 and 0.2 in Eq.~(\ref{eq:sea}). The first indications
for such a suppression came from early deep inelastic dimuon production
data, but now there is much firmer evidence for our
input assumption. The CCFR collaboration \cite{CCFR2} have performed
 a next-to-leading order  analysis of their  $\nu N \rightarrow \mu^-\mu^+X$
data and deduce that the strange sea distributions should lie within
the shaded band shown in Fig.~4. The strange sea that we find is
shown by the curve denoted by MRS(A) in Fig.~4 and satisfies the
experimental constraint.

The input charm sea is determined by the EMC deep inelastic data  for the
structure function $F_2^{c}$. We proceed as follows. We assume that
\begin{equation}
c(x,Q^2) = 0 \qquad \mbox{for} \qquad Q^2 < m_c^2
\end{equation}
and generate a non-zero distribution at higher $Q^2$ by massless GLAP evolution
at next-to-leading order. The structure functions are also calculated using
massless coefficient functions. Since $Q^2 = m_c^2$ falls below our
input scale, $Q_0^2 = 4\ \GeV^2$, we use an approximate set of partons to
evolve
between $m_c^2$ and $4\ \GeV^2$.
Taking zero charm at $Q^2 = m_c^2$ we find that the shape of the
resulting charm distribution generated at $Q^2 = 4\ \GeV^2$
is well described by the input parametrization
of the overall $S$ distribution. As might be expected the normalization,
specified by the parameter $\delta$ of (\ref{eq:sea}), depends sensitively on
the value
chosen for $m_c$. We adjust $\delta$ to give a good description of the
$F_2^{c}$ data for $Q \geq 5\ \GeV^2$. The fit is shown in Fig.~5 and
corresponds to $\delta = 0.02$, and to a charm distribution which can be
evolved from zero at $Q^2 = m_c^2 = 2.7\ \GeV^2$.
The value $\delta = 0.02$ implies that, at the the input scale
$Q_0^2 = 4\ \GeV^2$,
the charm sea carries $0.4\%$ of the proton's momentum, as compared to nearly
$4\%$ carried by the strange quark sea.

Our prescription  for the charm quark distribution is
only valid  far above threshold,
 $W^2 = Q^2(1-x)/x  \gg 4m_c^2$. Near threshold, a more
rigorous treatment of  quark mass effects is required.
As discussed in detail in Ref.~\cite{TUNG} (see also \cite{REYA}),
 various prescriptions are
possible. For example, one can {\it define} heavy quark densities
according  to our prescription, and absorb the threshold effects into
the coefficient functions. Nevertheless, Fig.~5 shows that
 our treatment of the charm
distribution does give a reasonable description of the EMC data
\cite{EMCCHARM}
for $F_2^{c}$ with $Q^2 > 5\ \GeV^2$. These data do in fact  lie in the region
$W^2 \gg 4 m_c^2$.  It is  interesting to note that our satisfactory
description
means that these EMC data show no necessity
for a small ``intrinsic" or non-perturbative  charm component, as advocated
by Brodsky {\it et al.} \cite{BRODSKY},
except possibly for one data point at $x=0.42$ (not shown) which lies well
above our fit.
 Also shown in Fig.~5 are
the predictions in the small $x$ region  which are relevant
for future measurements at HERA. Note that small $x$ implies
large $W^2$, and so our treatment should be reliable in this region.

 It has been argued \cite{BAR} that the
strange sea as measured in neutrino scattering should be different from that in
muon scattering on account of the different mass thresholds in $W^*g
\rightarrow
s\bar{c}$ as compared to $\gamma^*g \rightarrow s\bar{s}$.  In practice the
neutrino data have been corrected by the CCFR collaboration to approximately
take into account the $m_c \neq 0$ effects and to allow for this difference.
Our strange and charm seas should be interpreted as those  for muon data.
Our treatment of the charm sea at
large $x$ and modest $Q^2$ is suspect, but then
its contribution is too small to distort the analysis. Finally, the $b$-quark
contribution is included by assuming that
\begin{equation}
b(x,Q^2) = 0 \qquad \mbox{for} \qquad Q^2 < m_b^2
\end{equation}
with $m_b^2 = 30\ \GeV^2$, and non-zero contributions  at higher $Q^2$
generated by the same ``massless" prescription that was used for $c(x,Q^2)$.

The experimental data that are used to constrain the parton densities are
listed
in Table~1, together with the leading order partonic subprocesses, which helps
us to see which features of the distributions are constrained by the various
data sets.  As far as the global analysis is concerned, the \lq\lq core"
deep-inelastic data for $x \gapproxeq 0.01$ are the BCDMS \cite{BCDMS} and NMC
\cite{NMC1,NMC2} measurements of $F^{\mu p}_2$ and $F^{\mu n}_2$, and the CCFR
measurements \cite{CCFR1} of $F^{\nu N}_2$ and $xF^{\nu N}_3$.  It is useful to
inspect the leading order expressions of these structure functions in terms of
the quark densities
\begin{eqnarray}
\label{eq:f2pmn}
F^{\mu p}_2 - F^{\mu n}_2 & = & {\textstyle \frac{1}{3}}x(u + \bar{u} - d -
\bar{d}) \\
\label{eq:f2ppn}
{\textstyle \frac{1}{2}}(F^{\mu p}_2 + F^{\mu n}_2) & = & {\textstyle
\frac{5}{18}}x(u + \bar{u} + d + \bar{d} + {\textstyle \frac{4}{5}}s) \\
\label{eq:f2nu}
F^{\nu N}_2 = F^{\bar{\nu}N}_2 & = & x(u + \bar{u} + d + \bar{d} + 2s) \\
\label{eq:f3nu}
{\textstyle \frac{1}{2}}x(F^{\nu N}_3 + F^{\bar{\nu}N}_3) & = & x(u - \bar{u} +
d - \bar{d}) ,
\end{eqnarray}
where $N$ is an isoscalar nuclear target and where, for simplicity, we have
neglected the small $c$ quark contribution.  On a $(x,Q^2)$ point-by-point
basis, these four observables determine four combinations of quark densities,
which we may take to be $u + \bar{u}$, $d + \bar{d}$, $\bar{u} + \bar{d}$ and
$s$.  The difference $\bar{d}-\bar{u}$ is not determined and, moreover, at
leading-order it appears that the only constraint on the gluon is through the
momentum sum rule.

\begin{table}
\centering

\begin{tabular}{|l|l|l|}    \hline
                   &                           &                           \\
{\bf Process/}     &     {\bf Leading order}   & {\bf Parton determination}\\
{\bf Experiment}   &  {\bf subprocess}         &                           \\
                   &                           &                  \\  \hline
&\hfill \raisebox{-0.5ex}[0.5ex][0.5ex]{$\btop$}&                      \\
{\bf DIS} $\mbox{\boldmath $(\mu N \rightarrow \mu X)$}$ &  $\gamma^*q
\rightarrow q$ \hfill {\arrayrulewidth=1pt\vline}\hspace*{4pt}& Four structure
functions $\rightarrow$ \\
BCDMS, NMC& \hfill {\arrayrulewidth=1pt\vline}\hspace*{4pt}& \hspace*{1cm}
$u + \bar{u}$  \\
$F^{\mu p}_2,F^{\mu n}_2$& \hfill {\arrayrulewidth=1pt\vline}\hspace*{4pt}&
\hspace*{1cm} $d + \bar{d}$   \\
      &\hfill $\bmid$ & \hspace*{1cm}  $\bar{u} + \bar{d}$  \\
{\bf DIS} $\mbox{\boldmath $(\nu N \rightarrow \mu X)$}$ & $W^*q \rightarrow
q^{\prime}$  \hfill {\arrayrulewidth=1pt\vline}\hspace*{4pt}& \hspace*{1cm}
$s$ (assumed = $\bar{s}$),\\
CCFR (CDHSW)    &\hfill {\arrayrulewidth=1pt\vline}\hspace*{4pt}& but only
$\int xg(x)dx \simeq 0.5$ \\
$F^{\nu N}_2,xF^{\nu N}_3$    &\hfill {\arrayrulewidth=1pt\vline}\hspace*{4pt}&
[$\bar{u}-\bar{d}$ is not determined] \\
                &\hfill \raisebox{0.5ex}[1.5ex][1.5ex]{$\bbot$} &
                     \\
$\mbox{\boldmath $\mu N \rightarrow c \overline{c} X$}$ &
  $\gamma^*  c\rightarrow  c$    & $ c \approx 0.1 s$  at $Q_0^2$ \\
 $F_2^{c}$, EMC            &     &        \\
               &                            &                           \\
$\mbox{\boldmath $\nu N \rightarrow \mu^+\mu^-X$}$ &  $W^* s \rightarrow
c$    & $s \approx \frac{1}{2}\bar{u}$ (or $\frac{1}{2}\bar{d}$) \\
CCFR            &   $\;\;\;\;\;\;\;\;\;\;\;\;\;\hookrightarrow \mu^+$     &
\\
               &                            &                           \\
{\bf DIS (HERA})    &   $\gamma^*q \rightarrow q$   &   $\lambda$     \\
$F^{ep}_2$ (H1,ZEUS)   &                  & $(x\bar{q} \sim xg \sim
x^{-\lambda}$, via $g \rightarrow q\bar{q})$    \\
                      &                         &                     \\
$\mbox{\boldmath $p p \rightarrow \gamma X$}$  &  $qg \rightarrow \gamma q$  &
$g(x \approx 0.4)$ \\
WA70 (UA6)    &                  &                        \\
              &                   &                          \\
$\mbox{\boldmath $pN \rightarrow \mu^+\mu^- X$}$   &  $q\bar{q} \rightarrow
\gamma^*$  &  $\bar{q} = ...(1-x)^{\eta_S}$ \\
E605           &                     &                             \\
               &                     &                                 \\
$\mbox{\boldmath $pp, pn \rightarrow \mu^+\mu^- X$}$ & $u\bar{u},d\bar{d}
\rightarrow \gamma^*$   & $(\bar{u}-\bar{d})$ at $x = 0.18$  \\
NA51             &  $u\bar{d},d\bar{u} \rightarrow \gamma^*$   &     \\
                 &                   &                               \\
$\mbox{\boldmath $p\overline{p} \rightarrow WX(ZX)$}$    &  $ud \rightarrow W$
&  $u,d$
at $x_1x_2s \simeq M^2_W \rightarrow$  \\
UA2, CDF, D0          &             & \hspace*{1cm}  $x \approx 0.13$  CERN  \\
                       &             & \hspace*{1cm} $x \approx 0.05$ FNAL  \\
$\;\;\;\;\;\mbox{\boldmath $\rightarrow W^{\pm}$}$ {\bf asym}   &         &
slope of $u/d$ at $x \approx 0.05$  \\
$\;\;\;\;\;\;\;\;\;\;$ CDF            &              &              \\
                         &                         &          \\  \hline
\end{tabular}

\caption{The experimental data used to determine the MRS parton
distributions.  The last column gives an indication of the main type of
constraint imposed by a particular set of data.}

\end{table}

\bigskip

 From (\ref{eq:f2pmn}) and (\ref{eq:f2ppn}) we see that the strange quark
distribution is essentially
determined by the structure function difference
\begin{equation}
xs(x) \; \simeq \; {\textstyle \frac{5}{6}} F^{\nu N}_2(x) - 3F^{\mu D}_2(x) .
\label{eq:xs}
\end{equation}
At small $x \lapproxeq 0.08$ the value obtained is significantly larger than
that found in the dimuon analyses that we mentioned above.  We believe the
dimuon result is more reliable, as the determination based on
 (\ref{eq:xs}) is very  sensitive to the relative
normalization of two different data sets and to the heavy target corrections
that have to be applied to the neutrino data.  However it does mean that in the
global analysis this discrepancy will show up in the description of   $F^{\nu
N}_2$ and/or $F^{\mu D}_2$ at small $x$.  The quality of the MRS(A) fit to the
deep-inelastic data is shown in Figs.~6-8.  We note that the discrepancy occurs
in the description of $F^{\nu N}_2$ at $x \lapproxeq  0.05$, see Fig.~8(a).  We
allow an overall normalization parameter for the CCFR data.  When we fit to all
data (satisfying our $Q^2 > 5$ GeV$^2$, $W^2 > 10$ GeV$^2$ criteria) we find a
normalization factor of 0.95.  If, however, we were to exclude the $x \leq
0.08$
CCFR data then the normalization factor becomes 0.97 and the description of the
remaining neutrino data improves ($\chi^2 =$ 90/132 data points as compared to
171/160).

Heavy nuclear target corrections are applied to the neutrino data, and deuteron
screening corrections are made to the small $x$ muon-deuterium data as
described
in Ref.~\cite{MRSD1}.  As in the earlier analyses \cite{MRSD1,MRSD,MRSH} the
WA70 prompt photon data \cite{WA70} and the E605 Drell-Yan production data
\cite{E605} are included in the fit.  The former constrain the gluon and the
latter pin down the shape of the sea quark distributions.

The values of the parameters of the starting
 distributions, (\ref{eq:starting}), of the new MRS(A)
set of partons are listed in  Table~2.
 In addition, the value of the QCD scale
parameter is found to be  $\lmsb(n_f=4) = 230\ \MeV$, which corresponds to
$\alpha_s(M_Z^2) = 0.112_5$,
as it was for the MRS(H) set of partons.
 The parameter values for this latter set
are also listed in Table~2.
 Figure 9 shows the MRS(A) parton distributions as a function of $x$
for two different values of $Q^2$. A comparison of the ``new" MRS(A) and ``old"
MRS(H) partons is shown when we discuss the fit to the Drell-Yan asymmetry
measurement in section 4(a), since the differences between the two sets
arise mainly from introducing  this data point into the global analysis.
Finally, we show in Table~3  how the proton's momentum is shared
among the various parton flavours in the new MRS(A) set
 at different $Q^2$ values.

\begin{table}[t]
\centering
\begin{tabular}{|lccc|}
\hline
\rule[-1.2ex]{0mm}{4ex} &  &  MRS(A) & MRS(H) \\
\hline
 & $(A_g)$    & 0.775  & 0.777 \\
{\bf Glue} & $\lambda$ &  $-$0.3 & $-$0.3 \\
 & $\eta_g$ &   5.3 &  5.3 \\
 & $\gamma_g$ &   5.2   & 5.2 \\
\hline
 & $\eta_1$ & 0.538 &  0.335$^\dagger$ \\
 & $\eta_2$ &  3.96 & 3.90$^\dagger$ \\
 & $\epsilon_{u}$ &  $-0.39$ & 4.40$^\dagger$ \\
{\bf Valence} & $\gamma_{u}$ &  5.13 & 8.95$^\dagger$ \\
 & $\eta_3$ &  0.330 & 0.224 \\
 & $\eta_4$ &  4.71 &  4.65 \\
 & $\epsilon_d$ &  5.03 & 44.3 \\
 & $\gamma_d$ &   5.56 & 13.2 \\
\hline
 & $A_S$ &  0.411 & 0.386 \\
 & $\eta_S$ & 9.27 &  9.01 \\
{\bf Sea}  & $\epsilon_S$ &  $-1.15$ & 0.11 \\
 & $\gamma_S$ &  15.6 &  12.6 \\
 & $A_\Delta$ &  0.099 & 0.055 \\
 & $\gamma_\Delta$ &  25.0 &  --  \\
\hline
\end{tabular}
\caption{The numerical values of the
starting distributions (6)  of the  MRS(A)
set of partons.  For comparison we also list the values corresponding to
the MRS(H) partons\protect{\footnotemark}. Note that $A_g$ is fixed by the
momentum sum rule,
and is  therefore not a free parameter.}
\end{table}

\begin{table}[t]
\centering
\begin{tabular}{|c|rrrrrrrr|}
\hline
\rule[-1.2ex]{0mm}{4ex} $Q^2$ (GeV$^2$) & $g$ & $u_{\rm v}$ & $d_{\rm v}$ & $2
\bar u$ &
$2 \bar d$ & $2s$ & $2c$ & $2b$  \\
\hline
%\rule[-1.2ex]{0mm}{4ex}  $4$  &   &  &   &  &  &  &  & - \\
\rule[-1.2ex]{0mm}{4ex}  $7$  & 43.5  & 27.2 & 9.8 & 6.0 & 8.6 & 3.8 & 0.8 & -
\\
\rule[-1.2ex]{0mm}{4ex} $20$  & 44.8  & 25.1 & 9.1 & 6.3 & 8.6 & 4.2 & 1.4 & -
\\
\rule[-1.2ex]{0mm}{4ex}$100$  & 46.0  & 22.9 & 8.2 & 6.5 & 8.6 & 4.7 & 2.0 &
0.7 \\
\rule[-1.2ex]{0mm}{4ex}$10^4$ & 47.2  & 20.5 & 7.4 & 6.8 & 8.7 & 5.1 & 2.8 &
1.5 \\
\hline
\end{tabular}
\caption{Fractions (in \%) of the total proton momentum carried by the various
partons in the MRS(A) set.}
\end{table}
%      7. 0.2715 0.0978 0.0603 0.0855 0.0384 0.0075 0.0000 0.4352 0.9964
%     20. 0.2513 0.0905 0.0625 0.0858 0.0422 0.0136 0.0000 0.4477 0.9936
%    100. 0.2286 0.0824 0.0649 0.0861 0.0465 0.0204 0.0067 0.4599 0.9956
%   1000. 0.2053 0.0739 0.0676 0.0867 0.0511 0.0277 0.0155 0.4722 1.0000

\vspace*{1cm}

\noindent  {\large {\bf 3.  Small $x$ behaviour}}

The new measurements of $F^{ep}_2$ obtained by the ZEUS \cite{ZEUS} and H1
\cite{H1} collaborations in the low $x$ regime, $x \lapproxeq 0.005$, are shown
in Figs.~10 and 11.  These data were included in the global analysis and are,
in
fact, the only constraint on the parameter $\lambda$ in (\ref{eq:starting})
which controls the
small $x$ behaviour of the sea $xS \sim x^{-\lambda}$ (and of the gluon $xg
\sim
x^{-\lambda}$).  The existing set of partons, MRS(H) with $\lambda = 0.3$, is
found to give an excellent description of the new data, and so
it is not surprising  that  the new
parametrization, MRS(A), has the same value of $\lambda$.  The
description of the HERA data is shown in Figs.~10 and 11, together with that of
the extrapolations obtained from the pre-HERA
MRS(D$^{\prime}_0$,D$^{\prime}_-$)
parton sets.

\vspace*{.5cm}

\noindent  {\bf (a) Predictions from perturbative QCD}

The value of $\lambda$ is of great importance for understanding QCD dynamics in
the small $x$ regime.  Therefore before we discuss the uncertainty and the
implications of the $\lambda = 0.3$ determination, it is useful to briefly
review the perturbative QCD expectations for the small $x$ behaviour of $F_2$.
In the small $x$ regime we encounter large log$(1/x)$ contributions which have
to be resummed.  It is necessary to distinguish two different limits.  First,
the BFKL limit of small $x$ and moderate $Q^2 \gapproxeq Q^2_0$ in which we sum
the large log$(1/x)$ terms, but keep the full $Q^2$ dependence, not just the
leading log$Q^2$ contributions.  Second, the small $x$ and large $Q^2/Q^2_0$
region where the double-leading-logarithm approximation of GLAP evolution is
appropriate, in which the $\alpha_s$log$(1/x)$log$(Q^2/Q^2_0)$ contributions
are
resummed.  Ideally we seek a formalism which embodies both limits.  Some
progress has been made by Marchesini {\it et al.} \cite{MARCH} to find a
unified
evolution equation, but much remains to be done before we can extract the form
of the small $x$ behaviour of partons.  However we explain below why this is
not
a serious obstacle to the extraction of partons from the small $x$ HERA data.
\footnotetext{For MRS(H) and previous MRS analyses we parametrized
$x(u_{\rm v} + d_{\rm v})$  by the expression that we use here for
$x u_{\rm v} $. Thus parameters  marked $^\dagger$  in Table~2
correspond to $x(u_{\rm v} + d_{\rm v})$  and not to $x u_{\rm v} $.
To improve the precision at small $x$  we have repeated the MRS(H)
analysis and so the MRS(H) parameters listed in the Table are not
precisely the same as those of Ref.~\cite{MRSH}.}

First we review the BFKL expectations at small $x$.  The BFKL equation is
effectively the leading $\alpha_s$log$(1/x)$ resummation of soft gluon
emissions.  The equation may be solved numerically and the $k_T$-factorization
theorem used to predict the small $x$ behaviour of $F_2$.  It is found
\cite{AKMS} that
\begin{equation}
F_2(x,Q^2) \; = \; C(x,Q^2)x^{-\lambda_L} + F^{{\rm bg}}_2(x,Q^2)
\label{eq:f2decomp}
\end{equation}
where the coefficient $C$ of the BFKL contribution and the non-BFKL term,
$F^{{\rm bg}}_2$, are weakly dependent on $x$.  The {\it magnitude} of $C$ is
dependent on the treatment of the infrared region of the integration over the
transverse momenta of the emitted gluons -- but a physically reasonable choice
of the infrared parameter yields an $F_2(x,Q^2)$ in good agreement with the
HERA
data.  The important point is that the value of the exponent, $\lambda_L
\approx
0.5$, is a stable prediction.  That is, the $x^{-\frac{1}{2}}$ {\it shape} in
(\ref{eq:f2decomp}) is a characteristic property of (leading-order)
 BFKL dynamics; $\lambda_L$
is not a free parameter but is determined dynamically.
(For this reason we use the subscript $L$ to distinguish
the ``Lipatov" $\lambda_L$ from the free parameter $\lambda$
in Eq.~(\ref{eq:starting}).)  Of course the predicted
small $x$ behaviour may be modified by sub-leading corrections.  At
sufficiently
small $x$, shadowing corrections will almost certainly suppress the growth of
$F_2$ with decreasing $x$, but this effect is expected to be small in the HERA
regime, unless \lq\lq hot-spot" concentrations of gluons occur within the
proton
\cite{AKMS}.  The non-BFKL term, $F^{{\rm bg}}_2$, is one sub-leading
contribution which may be estimated
\begin{equation}
F^{{\rm bg}}_2(x,Q^2) \simeq \; F^{{\rm bg}}_2(x = 0.1,Q^2) \; \simeq \; 0.4,
\end{equation}
or perhaps rising slowly with decreasing $x$ as implied by the \lq\lq soft"
Pomeron.

We now turn to the second limit, that is the form of $F_2$ found at small $x$
and large $Q^2/Q^2_0$ from the double-leading-logarithm approximation of
conventional GLAP evolution.  In this case the result depends on the choice of
the exponent $\lambda$ in the input gluon and sea quark distributions
\begin{equation}
xg(x,Q^2_0), \; xS(x,Q^2_0) \; \sim \; x^{-\lambda}
\label{eq:singular}
\end{equation}
as $x \rightarrow 0$.  If {\it non-singular} input forms are used with $\lambda
\leq 0$, then
\begin{equation}
F_2(x,Q^2) \; \sim \; {\rm exp}(2[\xi
(Q^2_0,Q^2){\rm log}(1/x)]^{\frac{1}{2}}) ,
\label{eq:dllog}
\end{equation}
where the \lq\lq evolution length"
\begin{equation}
\xi (Q^2_0,Q^2) \; \equiv \; \int^{Q^2}_{Q^2_0} \frac{dq^2}{q^2}
\frac{3\alpha_s(q^2)}{\pi} .
\end{equation}
That is $F_2$ increases faster than any power of log$(1/x)$ but slower than any
power of $x$.  We see that the steepness in $x$ is not stable to evolution in
$Q^2$, but increases rapidly with the evolution length $\xi(Q^2_0,Q^2)$.  On
the
other hand if {\it singular} input forms, (\ref{eq:singular}), are chosen with
$\lambda > 0$,
then the $x^{-\lambda}$ shape
\begin{equation}
F_2(x,Q^2) \; \sim \; h(Q^2) x^{-\lambda}
\label{eq:bfkl}
\end{equation}
is stable to evolution in $Q^2$, and the behaviour (\ref{eq:bfkl}) overrides
the double
leading logarithmic structure of (\ref{eq:dllog}) \cite{JK}.

\vspace*{.5cm}

\noindent  {\bf (b) Implications of the global analysis in the HERA regime}

Can the HERA measurements of $F_2$ distinguish between the small $x$ behaviours
presented in (\ref{eq:f2decomp}), (\ref{eq:dllog})
and (\ref{eq:bfkl}), bearing in mind that the data are well
described with MRS(A) partons with a sea quark input form
\begin{equation}
xS(x,Q^2_0) \; \sim \; x^{-\lambda}(1 + \epsilon_S \sqrt{x} + ...)
\label{eq:xsea}
\end{equation}
with $\lambda  = 0.3$, and that $F_2$ mirrors the
behaviour of $xS$?

It is informative to first recall the GRV  dynamical parton model
of Gl\"uck, Reya and Vogt \cite{GRV}, which
pre-dated the HERA small $x$ measurements. GRV predicted a steep behaviour of
$F_2$, based on the double-leading-logarithm
form (\ref{eq:dllog}), by evolving from
valence-like input distributions at a very low scale, $Q^2_0 = 0.3$ GeV$^2$.
Although the GLAP small $x$ forms of
 (\ref{eq:dllog}) and (\ref{eq:bfkl}) are quite distinct, in a
limited $(x,Q^2)$ region about, say, $(\bar{x},\bar{Q}^2)$ the
double-leading-logarithm form mimics an $x^{-\bar{\lambda}}$ behaviour with
\begin{equation}
\bar{\lambda} \; = \; \left( \frac{36}{b_0} \frac{{\rm log}[{\rm
log}(\bar{Q}^2/\Lambda^2)/{\rm log}(Q^2_0/\Lambda^2)]}{{\rm log}(1/\bar{x})}
\right)^{\frac{1}{2}}
\end{equation}
where, for five flavours, $b_0 = 23$ and $\Lambda \approx 150$ MeV.  If we were
to take $Q^2_0 = 0.3$ GeV$^2$, as in the GRV model \cite{GRV}, then in the HERA
regime we have $\bar{\lambda} \gapproxeq 0.4$.  As we have seen, the HERA data
appear to prefer a slightly lower $\bar{\lambda}$.  We note that the observed
$x^{-0.3}$ behaviour can be approximately mimicked if we were to evolve from
non-singular input forms at a higher starting scale of $Q^2_0 = 1$ or 2
GeV$^2$,
see, for example, Ref.~\cite{KAI}.

The detailed $x$ and $Q^2$ behaviours of the GLAP-based forms,
(\ref{eq:dllog}) and (\ref{eq:bfkl}),
are quite distinct.  However if $Q^2_0$ is taken as a free parameter
 in (\ref{eq:dllog}) and
$\lambda$ as a free parameter in (\ref{eq:bfkl})
(or rather in (\ref{eq:xsea})), then the HERA data
are not yet precise enough to distinguish between them.  It will be even more
difficult to distinguish between the BFKL form (\ref{eq:f2decomp})
 and the GLAP-based behaviour
given by (\ref{eq:bfkl}).  In principle it might appear that the different
$Q^2$ dependence
of $C(x,Q^2) \sim (Q^2)^{\frac{1}{2}}$ in (\ref{eq:f2decomp}),
and of $h(Q^2) \sim {\rm log}Q^2$ in (\ref{eq:bfkl}),
would be a sufficient discriminator, but in practice the
differences are not large \cite{AGKMS}.

The most distinctive theoretical prediction of section (b) is the
$x^{-\lambda_L}$ behaviour with $\lambda_L \simeq 0.5$, embodied in
(\ref{eq:f2decomp}), which
was obtained from the leading-order BFKL equation.  Since $\lambda_L$ is not a
free parameter, the HERA data could have ruled out the precocious onset of this
BFKL behaviour.  Rather the data are found to be
consistent with (\ref{eq:f2decomp}), which may
be written in the form
\begin{equation}
F_2 \; \sim \; x^{-\frac{1}{2}}(1 + ax^{\frac{1}{2}})
\label{eq:f2phen}
\end{equation}
where $a = F^{{\rm bg}}_2/C$ is weakly dependent on $x$.  At $Q^2 \approx 15$
GeV$^2$ we estimate\footnote{The most recent AKMS \cite{AKMS} description,
(\protect{\ref{eq:f2decomp}}),
of HERA data (which can be found in Ref.~\cite{AGKMS}) has the approximate form
$F_2 = 0.018 x^{-\frac{1}{2}} + 0.385 (x/x_0)^{-0.08}$ at $Q^2 = 15$ GeV$^2$
where $x_0 = 0.1$.  Thus we have $a \equiv F^{{\rm bg}}_2/C \approx 0.6/0.018
\approx 30$ for $x$ in the HERA regime ($x \sim 5 \times 10^{-4}$).} that $a
\approx 30$.  Suppose we approximate the BFKL expectation,
(\ref{eq:f2phen}), by the simpler
form $F_2 \sim x^{-\lambda}$, then it follows that
\begin{equation}
\lambda \; \approx \; \frac{1}{2} -
\frac{ax^{\frac{1}{2}}}{2(1+ax^{\frac{1}{2}})} \; \approx \; 0.3
\end{equation}
if we insert  $a \approx 30$ and a typical value of $x$ in the HERA regime, $x
\sim 5 \times 10^{-4}$.  That is, the value of $\lambda$ predicted from BFKL
dynamics is in agreement with that found in the global analysis.  This very
approximate identification at $Q^2 = 15$ GeV$^2$ was simply made to illustrate
the similarity of the BFKL description with the global fit based on GLAP
evolution.  In fact present data for $F_2$ cannot distinguish between the two
descriptions.  An indication of the accuracy required to discriminate between
them can be found by comparing the MRS(H) and AKMS curves in Fig.~3 of Ref.~
\cite{AGKMS}.

To identify the BFKL behaviour (\ref{eq:f2phen}) with (\ref{eq:xsea})
 we have used the fact that the
$\epsilon_S $  term gives a small contribution for $x \sim
10^{-3}$ in the MRS(A) fit.  It will be no surprise to report that
in the global analysis there is a strong correlation between the values of the
$\lambda$ and $\epsilon_S$ parameters in (\ref{eq:xsea}).
 Almost as good a global
description of the data can be achieved with, for example,
\begin{equation}
\lambda = 0.2, \;\;\;  \epsilon_S = -3.3
\end{equation}
or with
\begin{equation}
\lambda = 0.4, \;\;\; \epsilon_S = 2.9
\end{equation}
In other words, the value of the parameter $\lambda$ in Eq.~(\ref{eq:starting})
 is not well determined by the HERA data.

It is important to note that it is the sea quark densities, and not the gluon,
which are constrained by the HERA measurements of $F_2$ at small $x$.
Since the sea quarks are driven by the gluon, via $g\to q \bar q$, we have
assumed that they have a common $x^{-\lambda}$ behaviour at small $x$, see
Eq.~(\ref{eq:starting}). However there is, as yet, no experimental confirmation
of this
assumption, and the ambiguity in the gluon distribution is by far the
largest uncertainty in the parton densities. It is therefore crucial
to make a direct determination of the gluon in the region $x \lapproxeq
0.05$. In section 5 we discuss the gluon distribution in more detail.

To summarize, we find that, within the global analysis, the HERA measurements
of
$F_2$ are well fitted by a parametrization which embodies a small $x$ behaviour
of the form $F_2 \sim xS \sim x^{-0.3}$.  This behaviour is consistent with the
expectations of BFKL dynamics, as expressed by $F_2 \sim Cx^{-0.5} + F_2^{{\rm
bg}}$ of (\ref{eq:f2decomp}), since this form mimics an $x^{-0.3}$ behaviour
in the HERA $x$
regime under consideration.  The observed $x^{-0.3}$ dependence is however not
consistent with the GRV \lq\lq valence" model which  gives a steeper behaviour
in the HERA region (with an effective $\lambda \gapproxeq 0.4$).  Such an
approach could be made consistent with the present data if a higher starting
scale $Q^2_0$ were used to decrease the evolution length.  We also noted that
very precise measurements of $F_2(x,Q^2)$ at HERA will be needed to distinguish
between the BFKL behaviour and GLAP evolution from \lq\lq singular" input
distributions (motivated by BFKL dynamics).  The converse of this result is of
great practical benefit to the determination of partons.  It means that we can
base the global analysis of the data, including the small $x$ measurements of
$F_2$, on GLAP evolution.  When precise data for $F_2$ become available over a
range of $x$ and $Q^2$ in the HERA regime, a global analysis may show a
systematic departure from the GLAP forms which would indicate BFKL dynamics
and/or the onset of shadowing corrections.

\vspace*{1cm}

\noindent {\large {\bf 4.  Constraints from the asymmetry measurements}}

As we saw in the introduction (Figs.~2 and 3) the existing set of partons
MRS(H), and its counterpart CTEQ2M, do not simultaneously describe the
Drell-Yan
and $W$ rapidity asymmetry measurements.  This is one of the major motivations
to redo the global analysis incorporating these data, and to present a new set
of partons -- MRS(A) -- which rectifies the deficiency.  The improvement is
shown by the MRS(A) curves in Figs.~2 and 3.

\vspace*{.5cm}

\noindent  {\bf (a) Drell-Yan asymmetry}

We have seen, from inspection of (\ref{eq:f2pmn})-(\ref{eq:f3nu}),
 that the ``core" deep-inelastic
structure function data offer little constraint on $\bar{d}-\bar{u}$.  An
indication that $\bar{d} \neq \bar{u}$ came from the NMC evaluation \cite{NMC}
of the Gottfried sum
\begin{equation}
I_{{\rm GS}} \; \equiv \; \int^1_0 \frac{dx}{x} (F^{\mu p}_2-F^{\mu n}_2) \; =
\; {\textstyle \frac{1}{3}} - I_{{\rm sea}} ,
\end{equation}
where the last equality separates the sum into a valence and a sea quark
contribution.  NMC directly measured the integral over the interval $0.004 < x
<
0.8$ and found $0.236 \pm 0.008$ (stat.) \cite{NMC}.  If we correct for
deuteron screening
effects and assume reasonable extrapolations over the unmeasured $x$ intervals
then the NMC result implies
\begin{equation}
I_{{\rm sea}} \; \equiv \; {\textstyle \frac{2}{3}} \int^1_0
(\bar{d}-\bar{u})dx
\; \approx \; 0.1
\label{eq:gsrsea}
\end{equation}
for $Q^2 \sim$ 4 to 7 GeV$^2$. This constraint on
$\bar{d}-\bar{u}$ given by (\ref{eq:gsrsea}),
which is sensitive to the
assumption made for the small-$x$ behaviour of the valence quarks, is not
directly included in the global analysis.  It is already correctly subsumed
since we fit to all the component NMC data.

The asymmetry in Drell-Yan production in $pp$ and $pn$ collisions offers a
direct determination of $\bar{d}-\bar{u}$ \cite{ES}.  The sensitivity can be
seen from the leading-order expression
\begin{equation}
A_{{\rm DY}} \equiv \frac{\sigma_{pp}-\sigma_{pn}}{\sigma_{pp}+\sigma_{pn}} =
\frac{(u-d)(\bar{u}-\bar{d}) +
\frac{3}{5}(u\bar{u}-d\bar{d})}{(u+d)(\bar{u}+\bar{d}) +
\frac{3}{5}(u\bar{u}-d\bar{d}) + \frac{4}{5}(s\bar{s}+4c\bar{c})}
\end{equation}
where $\sigma \equiv d^2\sigma/dMdy|_{y=0}$ and where the partons are to be
evaluated at $x = \sqrt{\tau} = M/\sqrt{s}$; $M$ and $y$ are the
invariant mass and rapidity of the produced lepton pair.
  Since $u > d$ in the proton, the
asymmetry is positive for parton sets with $\bar{d}-\bar{u}$ zero or small, but
becomes negative as $\bar{d}-\bar{u}$ increases.

When we include the recent NA51 asymmetry measurement \cite{NA51}, $A_{{\rm
DY}}
= -0.09 \pm 0.02 \pm 0.025$ at $x = 0.18$, in the next-to-leading
order analysis, the description of
the data is shown by the MRS(A) curve in Fig.~2.  Fig.~12 shows that the
introduction of the NA51 measurement leads to a significant increase in
$\bar{d}-\bar{u}$ in going from the \lq\lq old" MRS(H) to the \lq\lq new"
MRS(A)
partons.  In the region $0.02 \lapproxeq x \lapproxeq 0.7$, where a full set of
deep-inelastic data exist, we expect the MRS(A) partons to be very similar to
those of MRS(H), except that $\bar{d}-\bar{u}$ would be much larger while at
the
same time approximately conserving $\bar{u}+\bar{d}$, $d+\bar{d}$ and
$u+\bar{u}$ (as expected from (\ref{eq:f2pmn})-(\ref{eq:f3nu})).
Thus we anticipate an increase of
$\bar{d}$ which is compensated by a corresponding decrease in $\bar{u}$ and
$d$,
which in turn requires a similar increase in $u$.  The picture is a bit too
simplistic since there are new NMC data in the MRS(A) fit and moreover we have
to maintain the description of the $W$ rapidity asymmetry measurements.
Nevertheless the comparison of the MRS(A) and MRS(H) shown in Fig.~13 displays
the trends that we expect.

The NA51 collaboration also measure the asymmetry, $A_\psi$, of $J/\psi$
production in $pp$ and $pn$ collisions. Now $J/\psi$ production can be
of gg, as well as $q \bar q$, origin.
We may use a simple model \cite{MRS5}, in which
\begin{equation}
\hat{\sigma}(gg\to \psi) = r \hat{\sigma}(q \bar q \to \psi)
\end{equation}
where $r=0.5$, to estimate the asymmetry. This value of $r$ is obtained
 by comparing the rate of $J/\psi$ production in $pp$ and $p \bar p$ collisions
 \cite{MRS5}. To leading order we have
\begin{equation}
A_\psi = {\sigma_{pp}(\psi) - \sigma_{pn}(\psi) \over
\sigma_{pp}(\psi) + \sigma_{pn}(\psi)} =
{ (u-d)(\bar u - \bar d) \over  r gg + (u+d)(\bar u + \bar d)
+ 2 s \bar s + 2 c \bar c}  ,
\end{equation}
where $\sigma \equiv d \sigma/dy\vert_{y=0}$ and where the partons densities
are evaluated at $x = M_{\psi}/\sqrt{s}$ and $Q^2 = M_\psi^2$.
In principle it appears that a measurement of $A_\psi$ offers a more
direct determination of $(\bar u -\bar d)$ than $A_{DY}$, but in practice the
$gg$ term dominates and considerably dilutes the predicted asymmetry $A_\psi$.
As a consequence, the NA51 measurement \cite{NA51}
$ A_\psi\; =\; -0.03\; \pm\; 0.002\; {\rm (stat.)}\;
 \pm\; 0.02\; {\rm (sys.)}$ is not able to distinguish between the different
 sets of partons. However the value $A_\psi = -0.013$ predicted using the
 MRS(A) partons is  consistent with the measured value.

\vspace*{.5cm}

\noindent  {\bf (b) $W$ charge asymmetry}

The $W^\pm$ charge asymmetry at the FNAL $\pp$ collider
\begin{equation}
A_W(y) = { d\sigma(W^+)/dy -  d\sigma(W^-)/dy \over
 d\sigma(W^+)/dy +  d\sigma(W^-)/dy }
\end{equation}
is a sensitive probe of the difference between $u$ and $d$ quarks
in the $x \sim 0.1$, $Q\sim M_W$ region. Because the $u$ quarks carry more
momentum on average than the $d$ quarks,
the $W^+$ tend to follow the direction of the incoming
proton and the $W^-$ that of the antiproton, i.e. $A_W > 0$ for
$y > 0$. Thus a precise measurement
of the $W$ asymmetry -- in practice the asymmetry $A_\ell$ of the charged
lepton
{}from $W$ decay (\ref{lepasy}) -- serves as
 a valuable independent check on the $u$- and
$d$-quark distributions. In an earlier paper \cite{HMRS}, we showed that there
was a direct correlation between the lepton asymmetry and the {\it slope}
of the $d/u$  ratio. To see this, we first note that the dominant
contribution to $W^+\ (W^-)$ production comes from the $u\bar d\  (d\bar u)$
annihilation process. Thus
\begin{equation}
A_W(y) \simeq  {u(x_1) d(x_2) - d(x_1)u(x_2) \over u(x_1) d(x_2) +
d(x_1)u(x_2)}
\label{simpleasy}
\end{equation}
where the scale $Q = M_W$ is implicit for the parton distributions, and
\begin{equation}
x_{1,2} = x_0\; \exp( \pm y)\; ,\quad  x_{0} = {M_W\over \sqrt{s}}\; .
\end{equation}
If we introduce the ratio $R_{du}(x) = d(x)/u(x)$, then for small $y$
\begin{equation}
A_W(y) \simeq -\; x_0 \; y \ {R_{du}'(x_0)\over R_{du}(x_0) } \; .
\label{asyslope}
\end{equation}
In reality, the situation is of course more complicated -- it is
the {\it lepton} asymmetry  (\ref{lepasy}) which is
measured and there are subleading
and higher-order corrections to Eq.~(\ref{simpleasy}). Nevertheless,
the correlation implied by Eq.~(\ref{asyslope})  {\it is} evident
in the full prediction.\footnote{The curves in Fig.~3 are calculated using the
 next-to-leading order program DYRAD of Ref.~\cite{DYRAD}. We thank
 Nigel Glover for helping with these calculations of
 the $W$ asymmetry.}
Figure~14 shows the MRS(A), MRS(H) and CTEQ2M
$u$, $d$ and $d/u$ parton distributions
as a function of $x$ at $Q^2 = M_W^2$. The $x$ range is chosen
to correspond to the lepton asymmetry  measurement
by CDF at the FNAL $\pp$ collider \cite{BODEK}.
The slope  of $d/u$ is significantly larger  in
magnitude for the CTEQ set, and this leads to the larger  lepton asymmetry
shown  in Fig.~3.
MRS(H) partons give an excellent description of the $W$ asymmetry,
in fact slightly better (in terms of $\chi^2$) than MRS(A)
 which includes the data in the fit.
For this reason $u$ and $d$ do not change as much
  as might be expected to compensate for the increase in
$\bar d - \bar u$ in going from MRS(H) to MRS(A), see Fig.~13 and
the discussion following Eqs.~(\ref{eq:f2pmn})-(\ref{eq:f3nu}).

 From Fig.~3, we can conclude that CTEQ2M  is ruled out by the
data. This illustrates the discriminating power of the CDF data,
since CTEQ2M is consistent with all the deep inelastic
scattering data, and with the NA51 Drell-Yan asymmetry.
It is interesting to note that the CTEQ2M gives a reasonable
description of the NMC $F_2^n/F_2^p$ data \cite{NMC2} shown in Fig.~7,
a quantity which is also  correlated with the $d/u $ ratio.
The reason that there is no contradiction between this and the misfit
of the CDF asymmetry data is that the $n/p$ ratio is more influenced
by the $\bar u$ and $\bar d$ distributions than is $\sigma_W$ --
the former is, in a sense, {\it linear} in the small $\bar q/q$ correction,
$F_2^p \sim {\textstyle{4\over 9}} (u+d) + {\textstyle{4\over 9}} (
\bar u +\bar d)$, whereas the latter is {\it quadratic}:
$\sigma(W^+) \sim ud+ \bar d \bar u$.  Hence the $W$ asymmetry
is a more direct probe of the $d/u$ ratio than the $n/p$ structure
function ratio. The ``incorrect" $d/u$ ratio of CTEQ2M is compensated
by the antiquark contributions in the fit to the $n/p$ ratio.

Another place where the $d/u$ ratio is important is in the precision
$W$ mass determination in $\pp$ collisions. Because of the
finite rapidity acceptance of the experiments, the lepton transverse
momentum spectrum -- from which the mass is determined -- depends
on the shape of the rapidity distribution of the $W$ \cite{ADMWJS}.
Once again the correlation is with the $d/u$ ratio, and hence the $W$ asymmetry
can be used to rule out sets of parton distributions
in the $M_W$ analysis. In fact it should be possible to produce sets
of distributions which give ``$\pm 1 \sigma$" fits to the $W$ asymmetry
data of Fig.~3, and to use these to infer a $\pm 1\sigma$ error on $M_W$
{}from parton distributions.\footnote{We are grateful to Arie Bodek and
David Saltzberg  for discussions on this point.}  We shall return to this issue
 in a future study.

\vspace*{1cm}

\noindent {\large {\bf 5.  The gluon distribution}}

The new data which we have introduced into the
global analysis has enabled us to fine-tune the {\it quark }
distributions. In this section, we comment briefly on the status
of and prospects for information on the {\it gluon} distribution.

The only precise information on the gluon distribution is that the total
momentum fraction is about 43\% at $Q^2 = Q_0^2$ (Table~3).
Traditionally, information on the gluon {\it shape}
 has come directly from large $p_T$
prompt photon production in fixed-target $pN$ collisions, where
the $q g \to \gamma q$ process is dominant. In particular,
past and present MRS distributions have used the high precision WA70
data \cite{WA70} to constrain the shape of the gluon in the
$x \sim 0.3 - 0.4$ region.
Indirect information also comes from scaling violations in
fixed-target deep inelastic
scattering, where the gluon gives an important contribution to the $Q^2$
dependence particularly at small $x$. In previous studies \cite{MRSGLUE},
we have  investigated the interplay between the prompt photon and deep
inelastic data, and derived ``$\pm 1 \sigma$" gluons which attempted to span
the allowed range of the distribution at medium-to-large $x$.
We should stress that all our recent distributions, for example
MRS(D$_0'$, D$_-'$, H, A), are  global ``best fits'' to the data
(available at the time of the analysis) of all the processes listed in Table~1.
The HERA measurements of $F_2$ were not available
for the MRS(D)  analyses, but these data do not pin down
the gluon and so D$_0'$ and D$_-'$ can be used to demonstrate
the ambiguity in its behaviour. However the point we wish to
emphasize is that the recent global analyses are ``best fits'' to the
WA70 data, and so the spread of the gluons
obtained underestimates the uncertainty.
Nevertheless, it is  interesting to compare the variation
in the recent MRS gluon
distributions. Figure~15 shows the  D$_0'$, D$_-'$
and A gluons\footnote{The A and H gluons are essentially identical,
see Fig.~13.} at
$Q^2 = 20\ \GeV^2$.   We see immediately that there are regions in $x$ where
these are very different (note the logarithmic scale),
and regions where they are quite similar. We discuss
these in turn below.

\begin{itemize}

\item[{(i)}] At {\it very small} $x$, $x < O(0.01)$, the size and shape
of the gluon is controlled by the $x^{-\lambda}$ term in the
starting parametrization. Since the gluons
are constrained to be similar at larger $x$ (see below), a larger $\lambda$
implies a steeper, larger gluon in the very small $x$ region.
 As discussed in detail in section 3,
perturbative QCD (whether manifest in GLAP or BFKL dynamics)  always
ties the small $x$ behaviour of the sea quarks and gluons together,
 and in the MRS
fits we always use the same $\lambda$ parameter for both,
see Eq.~(\ref{eq:starting}).
Thus the spread of the  D$_0'$, D$_-'$ and A gluons in Fig.~15 is very similar
to that in the corresponding $F_2$'s at small $x$. Note, however,
that while we would expect  the ratio $\bar q/g$ to be approximately
constant as $x\to 0$, we cannot say at which value of $x$ this constancy
should set in. In fact, for MRS(A) this proportionality between the
distributions is valid over a broad range of small $x$, since the $\epsilon_S$
term turns out to be comparatively small in  the sea-quark distribution
and since there is no such parameter in the
gluon distribution\footnote{Our approach
has always been to use simple, physically-motivated forms for the starting
distributions and to only introduce extra parameters  as and when required
by the data.},
i.e. $\epsilon_g \equiv 0$ (Eq.~(\ref{eq:starting})).

Now that the very small $x$ quarks are being pinned down by the HERA data,
it is   important to measure the gluon at small $x$ to look for evidence
of a steeply rising distribution.
Various methods have been suggested
for directly extracting the gluon distribution  at HERA. In the long term,
the most promising appears to be the longitudinal structure function $F_L$
\cite{FL}. In the meantime, some very rough indication can be obtained
{}from the derivative $d F_2 / d \log Q^2$ which, like $F_L$, is dominated
by a term proportional to the gluon at small $x$. Figure~16 shows $xg$,
$F_L$ and $d F_2 / d \log Q^2$ at small $x$ for the
D$_0'$, D$_-'$ and A partons. Also shown are the first data on the $F_2$ slope
{}from H1 \cite{H1SLOPE}. The size of the errorbars
prevents any conclusions from being drawn at present.

 The production of large $p_T$ jet pairs
with large rapidity at the FNAL $\pp$ collider
 has also been suggested as a way of probing the very small $x$
 distribution \cite{CDFDIJET,MRSDIJET}. Preliminary indications appear
 to favour a steeply rising gluon, although a more complete study
 of the  experimental systematic errors and of the next-to-leading order
QCD corrections is required before any definite conclusions can be drawn.

\item[{(ii)}] At {\it medium-large} $x$,  $O(0.2) <x < O(0.4)$, the gluon
distributions  in Fig.~15 are constrained by the prompt photon data and are
therefore similar. This is {\it not} to say that there is no uncertainty
in the gluon this region -- in fact quite a large spread is allowed by the
errorbars on the data and by the factorization and scale dependence of the
next-to-leading order QCD cross section \cite{MRSGLUE}. Roughly
speaking, this translates into a $O(\pm 1)$ uncertainty on the power $\eta_g$
of $(1-x)$. Note that in the MRS parametrization
of the gluon, Eq.~(\ref{eq:starting}),
the parameter $\gamma_g$ compensates the different small $x$ behaviours
in, for example, D$_0'$, D$_-'$ and A, leading to roughly equal
distributions with the same parameter $\eta_g$ at larger $x$. The larger the
value of $\lambda$, the bigger is  the corresponding  value of $\gamma_g$.

\item[{(iii)}] At {\it  very large} $x$,  $x >  O(0.4)$, there are essentially
no constraints on the gluon distribution.  Phenomenologically, this
region is of little importance for high energy hadron-hadron colliders.
It is interesting, however, that the power of $(1-x)$ determined
by the prompt photon data at medium-large $x$, $\eta_g = 5.3$ for MRS(A)
at $Q_0^2 = 4\ \GeV^2$,
is perfectly consistent with the naive expectation from dimensional
counting, $\eta_g = 5$, which is supposed to be valid in the limit $x\to 1$
at some low $Q^2$ scale.

\item[{(iv)}] The {\it medium-small} $x$ region, $O(0.01) < x < O(0.2)$ is
perhaps the most interesting of all. Because of (a) the momentum sum rule
and (b) the requirement of similar gluons at larger $x$, and gluon
distribution which is much larger at very small $x$ must be smaller
in this region. This leads to a cross-over point at around
$x \sim 0.01$ for the gluons shown in Fig.~15.
One direct consequence of this  is that, in this region of $x$,
 structure functions such as $F_2$ evolve more slowly with $Q^2$ for
 sets with ``singular" gluons at very small $x$. This is illustrated
 in Fig.~17, which shows
 $F_2^p$ at $ x = 0.05$ as a function of
 $Q$, with data from NMC \cite{NMC1} and curves corresponding to
 the MRS(D$_0'$, D$_-'$, A, H) and CTEQ2M partons.
The fits are, of course, constrained to be in agreement in the
$Q$ range where accurate NMC data are available. Notice, however, the
divergence of the
predictions as $Q$ increases. This is directly related to the magnitude
of the gluons in this region of $x$ and $Q$, as is evident from Fig.~18 which
shows the
corresponding gluons at the same value of $x$, as a function of $Q$.
The effect of the gluon on the evolution of the quark distributions
at medium-small $x$ has important consequences for $W$ and $Z$ physics at the
FNAL $\pp$ collider. This will be discussed in more detail in the following
section.

There are, in fact, several hard scattering processes at the FNAL collider
which
offer, at least in principle, the possibility of measuring the gluon
distribution in the medium-small $x$ range. Data on prompt
photon production in the range $10 \ \GeV/c \lapproxeq p_T^\gamma \lapproxeq
 100 \ \GeV/c$
has recently become available \cite{CDFGAMMA}. Next-to-leading
order predictions using ``standard" gluon sets give a reasonable
description of the data for $p_T^\gamma \gapproxeq 30\ \GeV/c$ \cite{CDFGAMMA}.
Below this value, the measured cross section rises more rapidly
than the predictions, indicating an  excess of data over theory.
However this is precisely the region where (a) the uncertainties due
to scale dependence and  to
   matching the theoretical
and experimental definitions of
``isolated"  photons are greatest, and (b) one expects significant
contributions from higher-order processes. It therefore it seems
 premature to conclude \cite{REYAGAMMA} that the data indicate
 a steeply-rising gluon in this $x$ region.
Artifically enhancing the gluon  to fit the CDF low
$p_T^\gamma$ data, at the same time keeping the overall
 gluon momentum fraction fixed, inevitably depletes the large $x$
 part of the distribution, thus
 spoiling the fit to the WA70 data.\footnote{We are grateful to
 Steve Kuhlmann for discussions on this point.}

The $b \bar b$ cross section, being proportional to the square of the gluon
distribution, is also of potential importance. Again, unfortunately,
there are substantial theoretical uncertainties arising  from  a large
scale dependence, and experimental uncertainties in reconstructing the
total cross section from the measured distributions of $B$ mesons, leptons,
$J/\psi$'s  etc. in restricted kinematic ranges, see for example
Ref.~\cite{MANGANO}. Also $b \bar b$ production at the FNAL $\pp$ collider
samples $x$ values ($ x \lapproxeq 0.01$) just below the
cross-over point where the various gluons have similar
magnitudes, see Fig.~15. The most recent
indications  \cite{FNALBB} are that while there is no serious
disagreement between theory and experiment, the uncertainties are still
much too large to allow any discrimination between gluon distributions
at the level required by  Fig.~15.

\end{itemize}

In summary, there is still a considerable amount of uncertainty
in the gluon distribution, particularly in the important
 medium-small and very small
$x$ regions. HERA will eventually provide information on the latter,
and in the meantime large rapidity jet  and prompt photon  production
at the FNAL $\pp$ collider are beginning to provide useful constraints.
Current studies tend to use sets like MRS(D$_0'$) and  MRS(D$_-'$)
to quantify the effect of different small-$x$
gluon shapes on  cross section predictions,
even though the corresponding quark  distributions are now ruled out by the
HERA $F_2$ data. In a future study, we will explore the possibilities
for varying  the gluon distribution while maintaining a good fit to the
structure function data.

\vspace*{1cm}

\noindent {\large {\bf 6. Hadroproduction of $W$ bosons and top quarks}}

In this section we present  predictions for the total
$W$ and $t \bar t$ cross sections in $\pp$ collisions at $1.8\ \TeV$.
The former is being measured more and more precisely by both CDF and D0,
and already provides a valuable cross check on the size of the
$u$ and $d$ distributions. Interest in the latter  has recently
been rekindled by the evidence for top production presented by the
CDF collaboration  \cite{CDFTOP}, with the suggestion
 that the measured production cross
section is larger than the theoretical prediction.

\vspace*{.5cm}

\noindent  {\bf (a) $W$ production cross section}

While the $W$ asymmetry probes the relative size of the $u$ and $d$
distributions, the
 {\it total} cross sections for $W$ and $Z$ boson production
in $\pp$ collisions at $\sqrt{s} = 1.8\ \TeV$ provide an important
check of the overall magnitude of the
quark distributions in a region of $x$ where they are constrained
by deep inelastic (mainly NMC and CCFR) data.
Since the subprocess cross section is known
to next-to-next-to-leading order \cite{LEIDEN}, there is very little
theoretical
uncertainty in the predictions once the  parton distributions are
specified.

To illustrate this, we show in Fig.~19(a) the total $W$ production
cross section
for the MRS(A,H,D$_0'$,D$_-'$) and CTEQ2M  parton distributions.
 The  central points are
calculated assuming $M_W = 80.21 \ \GeV/c^2$ and setting
the renormalization and factorization scales equal to $M_W$. The
errorbars  show the effect of changing both
scales simultaneously by a factor of $2$.
The first point to note is that the $\sigma_W$ predictions for
MRS(A), MRS(H) and CTEQ2M are very similar.
 (Notice also  that the scale dependence is weak, and gives a similar variation
 in cross section as  the spread in the three predictions.)
This can be understood by recalling that $\sigma_W$
is largely determined by the product $u(x_W,M_W^2)
 d(x_W,M_W^2)$ where $x_W = M_W/\sqrt{s} \sim 0.05$. By comparing
 the relative size of the $u$ and $d$ distributions for the MRS(A), MRS(H)
 and CTEQ2M sets, Fig.~14, we can understand the similarity of the
 corresponding
 $\sigma_W$ predictions shown in Fig.~19(a). First, we have that
\begin{equation}
\left.  \begin{array}{r}
 u({\rm A}) > u({\rm H})  \\
 d({\rm A}) < d({\rm H})
\end{array} \right\}
\quad\Rightarrow\quad\sigma_W({\rm A}) \simeq \sigma_W({\rm H}) \; .
\end{equation}
The situation with CTEQ2M is more subtle. By comparing the
$u$ and $d$ distributions at $x_W$, we would conclude that
\begin{equation}
\left.  \begin{array}{r}
 u({\rm A}) \simeq  u(2{\rm M})  \\
 d({\rm A}) > d(2{\rm M})
\end{array} \right\}
\quad\Rightarrow\quad\sigma_W({\rm A}) > \sigma_W(2{\rm M}) \; ,
\end{equation}
whereas in fact the reverse is true. The reason lies in the
different shapes of the $W$ rapidity distributions. The CTEQ2M  $W$
cross section receives a significant contribution from $W^+$ ($W^-$)
produced at large positive (negative) rapidity, where a large-$x$
$u$-quark interacts with a small-$x$ $d$-quark.  Especially for the former,
the CTEQ2M distribution is significantly larger than the MRS(A)
distribution, see Fig.~14, giving a distinctly different $W$
rapidity distribution, as shown in Fig.~20 for $p \bar p \to W^+ + X$.
 This is, of course,
the origin  of the much too large CTEQ2M  lepton  rapidity asymmetry discussed
in
section~4(b) and shown in Fig.~3. Comparing MRS(A) and MRS(H), we see that the
distribution in $y_W$ for the latter is
slightly larger for $y_W < 0$, which is reflected in the total cross section
shown in Fig.~19(a).
Since the $\lmsb$ values and gluon distributions (for $x \sim x_W$,
see Fig.~18) for these three sets are very similar, the above
orderings of $u$ and $d$ at $Q = M_W$ simply reflect the orderings
at the lower scales where the distributions are constrained by
the fixed-target deep inelastic  and asymmetry data discussed in the
previous sections.
Of the three sets only MRS(A) gives a
consistent fit to {\it all} the data,  and so we may conclude
 that $\sigma_W({\rm A})$
is the most accurate prediction.

It is also worth commenting on the differences between the MRS(H),
MRS(D$_0'$) and MRS(D$_-'$) $W$ cross sections. The three  parton sets
have similar $u$- and $d$-quark distributions for $x \sim x_W$
and $Q^2$ values characteristic of the fixed target deep inelastic scattering
data \cite{MRSH}. There is, however, a rather large spread in  the
 corresponding $W$ cross sections  shown in Fig.~19(a).
These differences arise from the different  {\it gluon} distributions
in the three sets.
We have seen in the previous
section that $x \sim 0.05$ is precisely the region where
the gluon is not well constrained. Thus even though the quarks are
accurately determined at low $Q$ for this $x$ range, differences
in the gluon can affect the GLAP evolution to $Q = M_W$ and lead  to
differences in the quarks at this scale. This is graphically
illustrated by the $\sigma_W(H)$,
$\sigma_W({\rm D}_0')$ and $\sigma_W({\rm D}_-')$
predictions shown in Fig.~19(a). From Fig.~18, we see that
the ordering of the gluons at low $Q$ and $ x = 0.05$ is
$ g({\rm D}_-') < g({\rm H}) < g({\rm D}_0')$, and hence the quarks of
D$_-'$/D$_0'$ increase more slowly/rapidly than those of  H.
This leads directly to $\sigma_W({\rm D}_-') < \sigma_W({\rm H})
< \sigma_W({\rm D}_0')$. It would appear, therefore, that now
that the quarks are very well constrained, it is the
uncertainty in the gluon  distribution around $x \sim 0.05$
which constitutes the largest theoretical uncertainty in the predictions
for the $W$ (and $Z$) cross sections at the FNAL $\pp$ collider.

A precision measurement of $\sigma_W$ at the percent level
 would  obviously  provide a useful additional constraint on the
distributions.
The most recent  values \cite{CDFSIGW,D0SIGW} for $\sigma_W$
are\footnote{We have  divided out the Standard Model leptonic
branching ratio $B(W\to\l\nu) = 0.108$.}
\begin{eqnarray}
\mbox{CDF}(e,\mu)\ :  & &  \sigma_W = 20.4  \pm 0.4\;{\rm (stat)}\;
\pm 1.2\;{\rm (sys)} \;  \pm 1.4\;{\rm (lum)} \ \nb \nonumber \\
\mbox{D0}(e,\mu)\ :  & &  \sigma_W = 19.0  \pm 0.6 \; {\rm (stat + sys)}
 \;  \pm 2.3\;{\rm (lum)} \ \nb
\label{wcross}
\end{eqnarray}
We  should
note that these cross sections relied on a  luminosity measurement
which has since been superceded by a new measurement of the
total $\pp$ cross section by CDF \cite{CDFTOTAL}.  If the new values
are indeed some 10\% higher, as suggested by the ratio of `new' and `old'
total  cross section measurements, then this would bring
 them more into line
with the theoretical predictions shown in Fig.~19(a).
However the errors on the data in Eq.~(\ref{wcross}) are still much too big
to provide the level of
discrimination between the distributions discussed above.
Indeed the accuracy of the theoretical prediction suggests that the $W$
cross section could itself serve as a determination of the collider
luminosity.

Finally, we note that HERA will eventually be able to make a reasonably
precise measurement of $F_2^p$ in a region of $Q$ intermediate
between the fixed target data and $M_W$ (see Fig.~17).
In fact the first data in this
kinematic region have recently been presented by H1 \cite{H1} and ZEUS
\cite{ZEUS}, although the errorbars are so large  that no meaningful
discrimination between the predictions is possible at present.

\vspace*{.5cm}

\noindent  {\bf (b) $t \bar t$ production cross section}

The evidence presented recently by the CDF
collaboration \cite{CDFTOP}  has focussed attention
on the theoretical prediction for  the top quark
production cross section. Here we address the question of the
uncertainty in the calculation of  $\sigma_{\tt}$ coming
{}from parton distributions. Fig.~19(b) shows the next-to-leading
order QCD predictions \cite{TOPNLO}
for the total cross section for $m_t = 174
\ \GeV/c^2$ -- the central value reported by CDF -- with the
factorization and renormalization scales set equal to $m_t$.
The errorbars on  the  predictions indicate the effect of changing
the scales by a factor of 2.
The relative spread in the predictions is somewhat less than for
$\sigma_W$, since now the quark distributions are being probed
in the range of $x \sim 0.2 - 0.4$ where the high-statistics
BCDMS and CCFR data provide a very tight constraint.\footnote{Note that the
$q \bar q$ annihilation process accounts for approximately 90\% of
the total cross section for this value of $m_t$.}
The more important point, however, is that unlike  for $\sigma_W$,
the dominant effect is the scale dependence.   As discussed in detail
in Ref.~\cite{LAENEN}, this scale dependence
is symptomatic of large contributions
to the cross section from soft gluons at higher orders in perturbation theory.
Fig.~21 shows the next-to-leading
order prediction for $\sigma_{t \bar t}$ as a function of $m_t$
using the MRS(A) partons. The solid curve corresponds to the scale choice
$Q=m_t$ and the dashed curves to $Q=2 m_t, {\textstyle{1\over 2}} m_t$.
 The data point is
the CDF measurement \cite{CDFTOP}, $\sigma_{t \bar t} =
13.9 {+ 6.1\atop -4.8} \ \pb$ for
$m_t = 174 \pm 17\ \GeV/c^2$.

\vspace*{1cm}

\noindent {\large {\bf 7. Conclusions}}

To determine the partonic structure of the proton we have performed
a global next-to-leading order analysis of the data for deep inelastic
scattering and related hard-scattering processes. We summarized the
constraints from the various types of data in Table~1. We found that a
surprisingly economical parametrization of the parton densities at $Q^2
= 4\ \GeV^2$ was able to give a satisfactory description of the data. We
were able to include in the analysis for the first time significant new
measurements of $F_2$ at HERA, of the asymmetry $A_{DY}$ in
Drell-Yan lepton pair production from $pp$ and $pn$ collisions,
and of the $W^\pm$ rapidity asymmetry $A_W$ at FNAL. The partons that we
obtain -- MRS(A) -- are the only set which are consistent with all the
data. (The only exception which remains is the incompatibility with the
lowest $x$ measurements by CCFR of the neutrino structure functions $F_2$ and
$xF_3$.)

The asymmetry data considerably tighten the constraints on the $u$, $d$,
$\bar u$ and $\bar d$ distributions. In particular, the NA51 measurement
of $A_{DY}$ provides a missing link and pins down the combination $\bar
d - \bar u$, which is essentially undetermined by the deep-inelastic
structure function measurements. The measurements of the $W^\pm$
lepton rapidity asymmetry,  $A_W$, provide a tight constraint on the
ratio $d/u$ with $\bar u$ and $\bar d$ having less influence.

An interesting feature to emerge from the analysis is the link between
the $u$ and $d$ distributions at $Q^2 \sim M_W^2$  and $x \sim 0.05$,
which determine the $W$ cross section at FNAL, and the $u$ and $d$
densities determined from the fixed-target data with $Q^2 \sim 20\
\GeV^2$. Even for $x$ as large as $0.05$, the evolution from $Q^2 \sim
20\ \GeV^2$ to $M_W^2$ depends appreciably on the gluon distribution,
particularly in the initial stages of the evolution.

The measurements of $F_2$ at HERA have opened up the small $x$ regime,
$x \lapproxeq 10^{-3}$. The dramatic growth of $F_2$ observed with
decreasing $x$, which had been anticipated in perturbative QCD models,
is readily accommodated in the global fit. Indeed we found that the
observed small $x$ behaviour is entirely consistent with the precocious
onset of BFKL leading $\log(1/x)$ dynamics, although alternative QCD
explanations are not ruled out.

We believe that our analysis has considerably improved the detailed
knowledge of the quark densities. However, the same is not true for the
gluon distribution. At present there are, with the possible exception of
the WA70 prompt photon measurements, no reliable direct constraints on
the gluon. In principle, $b \bar b$ and jet production cross sections
depend directly on the gluon density, but ambiguities -- due to scale
dependence, jet recognition, higher-order effects, the choice
of $m_b$ etc. -- mean that so far these data have provided little
additional information on the gluon.     Valuable indirect constraints
on the gluon come from the momentum sum rule and from the observed
structure of the deep-inelastic scaling violations.  Our philosophy has
been to keep the parametrization of  the partons as economical as
possible, and to only introduce an extra parameter when it is required
to improve the description of the data.  For this reason the
parametrization of the gluon is remarkably simple, such that its
behaviour is prescribed by the momentum sum rule and the optimum
fit to the WA70 prompt photon data.   It is therefore not surprising
that the gluons from the various sets of partons are quite similar above
$x \sim 10^{-2}$. Clearly this does not reflect the true ambiguity in
the gluon distribution, but rather it is an artefact of our economical
parametrization. We will attempt to address this problem in a future
publication. We will also evolve our distributions to lower $Q^2$,
below $Q_0^2 = 4\ \GeV^2$, and study the interplay of leading and
higher twists, and hence provide a phenomenological representation
of $F_2$ at  low $Q^2$.

For the future, we foresee a continued improvement in our knowledge of
the partons. The increasingly precise measurements of $F_2$ at HERA,
particularly at low $x$ and low $Q^2$ ($Q^2 \sim 5\ \GeV^2$) may help to
distinguish between BFKL and GLAP dynamics and may even reveal the onset
of shadowing behaviour. So far the Drell-Yan asymmetry has only been
measured at $x = 0.18$. This leaves considerable ambiguity in the
structure of $\bar d - \bar u$. The results of the asymmetry measurement
by a forthcoming experiment at FNAL \cite{GARVEY} will cover an extended
$x$ interval and are eagerly awaited. The accuracy of the $W^\pm$
rapidity asymmetry  measurements at FNAL will improve and thus provide
extremely tight constraints on the behaviour of $d/u$ in the $x$ region
around $0.05$. The neutrino measurements at low $x$ are under further
study by the CCFR collaboration and the discrepancy with the NMC
measurements may be resolved. A change may effect the relative
normalization between the data sets -- an improvement here would be of
great value. Finally,  detailed studies of $J/\psi$, $c \bar c$,
$b \bar b$, prompt photon
and jet production at FNAL and HERA, together with measurements
of the longitudinal  structure function  $F_L$, will help reduce the
ambiguity in the gluon distribution.

\vspace*{1cm}

\noindent {\large {\bf Acknowledgements}}

We are grateful to Hiro Aihara, Arie Bodek,
 Antje Bruell, Robin Devenish, Albert De Roeck, Marcel Demarteau,
 Mark Dickson, Kevin Einsweiler,
 Henry Frisch, Walter Giele, Nigel Glover, Bill Gregory,
 Louis Kluberg, Sacha Kopp, Michelangelo Mangano,
Jeff Owens, Ewa Rondio and Paul Slattery for useful discussions.

\newpage
\section*{Figure Captions}
\begin{itemize}
\item [{[1]}]
The H1 and ZEUS data \cite{HERA} obtained from the 1992
HERA run, but published in 1993, together with earlier
data (interpolated to $Q^2=15 \ \GeV^2$) obtained by the
NMC \cite{NMC1} and BCDMS collaborations \cite{BCDMS}.
The ``pre-HERA" extrapolations obtained using
MRS(D$^{\prime}_0$,D$^{\prime}_-$) parton sets \cite{MRSD}
were expected to span the forthcoming HERA measurements.
The ``post-HERA" curve H is the result of a global analysis
\cite{MRSH} which included the HERA data.

\item [{[2]}]
The measurement of the asymmetry in Drell-Yan production in
$pp$ and $pn$ collisions made by the NA51 collaboration \cite{NA51}
at $x=\sqrt{\tau}=0.18$. The curves obtained from MRS(H) \cite{MRSH}
and CTEQ2M \cite{CTEQ} partons pre-date the measurement.  The MRS(A)
curve is obtained from the global fit, presented in this paper,
which includes the NA51 data point.

\item [{[3]}]
The asymmetry $A(y_{\ell})$ of the rapidity distributions of the
charged leptons from $W^{\pm} \rightarrow \ell^{\pm}\nu$ decays
observed at FNAL \cite{BODEK} as a function of the lepton rapidity
$y_{\ell}$.  The curves are the next-to-leading order descriptions
obtained using MRS(H) \cite{MRSH}, CTEQ2M \cite{CTEQ}  and the
new MRS(A) partons.
The MRS(A) analysis, presented in this paper, includes the
data in the global fit.

\item [{[4]}]
The shaded band is the strange sea quark distribution,
$xs(x,Q^2=4\ \GeV^2)$, determined by the CCFR collaboration
\cite{CCFR2} from a next-to-leading order analysis of their
dimuon production data.  In addition they find that  $s/(\bar{u} +
\bar{d})$
is, to a good approximation, independent of $x$.  Also shown is
the MRS(A) input strange sea quark distribution, $xs(x,Q_0^2=4\
\GeV^2)$, with the 50\% suppression, as given in Eq.~(7).

\item [{[5]}]
The description of the EMC data \cite{EMCCHARM}
for  $F_2^{c}$ by the MRS(A) partons. We assume that the
$c \rightarrow \mu + X$ branching ratio is 8\%.

\item [{[6]}]
The description of the BCDMS \cite{BCDMS} and NMC \cite{NMC1}
measurements of the $F^{\mu p}_2 (x,Q^2)$ structure function by
the MRS(A) set of partons. The BCDMS data are shown with an overall
renormalization by a factor of 0.98.

\item [{[7]}]
The continuous curves show the description of the NMC data \cite{NMC2}
for the structure function ratio  $F^{\mu n}_2 /F^{\mu p}_2$
given by the MRS(A) set of partons.  The dotted curves are the predictions
of the MRS(H) partons obtained from a global analysis which
included an earlier set of NMC data for $F^{\mu n}_2 /F^{\mu p}_2$.

\item [{[8]}]
The MRS(A) fit to the CCFR measurements \cite{CCFR1} of the
structure functions $F^{\nu N}_2$ and $xF^{\nu N}_3$.  The data are
shown after correction for the heavy target effects and after an
overall renormalization of 0.95 required by the global analysis,
see text. The statistical, systematic and heavy target correction
errors have been combined in quadrature, see also Ref.~\cite{MRSD}.

\item [{[9]}]
The MRS(A) partons shown as a function of $x$ at $Q^2=10$ and
$10^4\ \GeV^2$.

\item [{[10]}]
The description of the preliminary ZEUS measurements \cite{ZEUS}
of $F^{ep}_2$ for $x<0.005$ by the parton sets of Refs.~\cite
{MRSD,MRSH}.  The long dashed curves show the fit to the data
obtained by the MRS(A) analysis presented in the paper.

\item [{[11]}]
As for Fig.~10 but showing the preliminary H1 measurements \cite{H1}.

\item [{[12]}]
$x(\bar{d}-\bar{u})$, as a function of $x$,
at $Q^2=7\ \GeV^2$ obtained from the
MRS(H) \cite{MRSH}, CTEQ2M \cite{CTEQ} and the new MRS(A) sets
of partons.

\item [{[13]}]
A comparision at $Q^2=20\ {\GeV}^2$
of the new MRS(A) partons of this analysis and the
MRS(H) partons of Ref.~\cite{MRSH}.

\item [{[14]}]
The $u$, $d$ and $d/u$ distributions at $Q^2 = M_W^2$  obtained from the
MRS(H) \cite{MRSH}, CTEQ2M \cite{CTEQ} and the new MRS(A) sets
of partons.

\item [{[15]}]
The gluon distributions from sets D$_0'$ (dotted), D$_-'$ (dashed) and A
(continuous curve) at $Q^2 = 20\ \GeV^2$.

\item [{[16]}]
 The gluon distributions  and the corresponding predictions for
  $d F_2 /d \log Q^2$ and $F_L$ for
 the D$_0'$, D$_-'$ (dashed lines) and A (solid line) parton
 distributions, together with data from H1 \cite{H1SLOPE}.

\item [{[17]}]
The structure function $F_2^p$ at $ x = 0.05$ as a function of
 $Q$, with data from NMC \cite{NMC1} and the descriptions
of the MRS(D$_0'$, D$_-'$, A, H) and CTEQ2M partons.

\item [{[18]}]
The gluon distributions $x g(x,Q^2)$ of  the  MRS(D$_0'$,D$_-'$,A,H)
and CTEQ2M partons,  at $ x = 0.05$, as a function of  $Q$.

\item [{[19]}]
(a) Predictions for $\sigma_W$ and
(b) $\sigma_{t \bar t}\ (m_t = 174\ {\rm GeV}/c^2)$
in $\pp$ collisions at $\sqrt{s} = 1.8\ \TeV$, from the
MRS(D$_0'$, D$_-'$, A, H) and CTEQ2M partons.
The central values are calculated using renormalization and
factorizations scales equal to (a) $M_W$ and (b) $m_t$, and the
errorbars indicate the effect of increasing and decreasing these
by a factor of 2.

\item [{[20]}]
Next-to-leading order predictions
 for  the $W^+$ rapidity distribution in $\pp$ collisions
 at $\sqrt{s} = 1.8\ \TeV$ from the
MRS(A, H) and CTEQ2M partons.

\item [{[21]}]
Next-to-leading order predictions
 for $\sigma_{t \bar t}$ at $\sqrt{s} = 1.8\ \TeV$ as a function of  $m_t$
using  MRS(A) partons. The curve is calculated using renormalization
factorization scales equal to $m_t$, and the
 band corresponds to changing scales to
${\textstyle{1\over 2}} m_t$ and $2 m_t$.
The data point is from the CDF collaboration \cite{CDFTOP}.

\end{itemize}

\end{document}